\documentclass[twocolumn,aps,floatfix,showpacs,prd,longbibliography]{revtex4-2}
\usepackage{graphics, epsfig ,color} %
\usepackage{amsmath,amssymb} %
\usepackage{amsfonts,subfigure}
\usepackage{ulem}
\usepackage{enumerate}
\usepackage[dvipsnames]{xcolor}

\usepackage{xcolor, soul}
\usepackage{todonotes}

\sethlcolor{yellow}

\graphicspath{{figures/}}

\begin{document}
\date{\today}
\title{Generalization emerges from local optimization in a self-organized learning network}
\author{S. Barland, L. Gil }
\affiliation{Universit\'e C\^ote d'Azur, Institut de Physique de Nice (INPHYNI), France}

\begin{abstract}
We design and analyze a new paradigm for building supervised learning networks, driven only by local optimization rules without relying on a global error function. Traditional neural networks with a fixed topology are made up of identical nodes and derive their expressiveness from an appropriate adjustment of connection weights. In contrast, our network stores new knowledge in the nodes accurately and instantaneously, in the form of a lookup table. Only then is some of this information structured and incorporated into the network geometry. The training error is initially zero by construction and remains so throughout the network topology transformation phase.  The latter involves a small number of local topological transformations, such as splitting or merging of nodes and adding binary connections between them. The choice of operations to be carried out is only driven by optimization of expressivity at the local scale. What we're primarily looking for in a learning network is its ability to generalize, i.e. its capacity to correctly answer questions for which it has never learned the answers. We show on numerous examples of classification tasks that the networks generated by our algorithm systematically reach such a state of perfect generalization when the number of learned examples becomes sufficiently large.  We report on the dynamics of the change of state and show that it is abrupt and has the distinctive characteristics of a first order phase transition, a phenomenon already observed for traditional learning networks and known as \textit{grokking}. In addition to proposing a non-potential approach for the construction of learning networks, our algorithm makes it possible to rethink the grokking transition in a new light, under which acquisition of training data and topological structuring of data are completely decoupled phenomena.
\end{abstract}
 \pacs{Self organized systems, Phase transitions, Physics of computation, Machine Learning Models}
\maketitle

\section{Introduction}

The computational properties of complex systems have long been a topic of fundamental interest for the physics community, from early examples of neural networks and cellular automata \cite{hopfield1982neural,langton1990computation} to very recent approaches of physics-based neuromorphic computing \cite{markovic2020physics}. A common task for such systems is that of \textit{supervised learning}, in which a neural network is trained to realize a particular function based on many examples of input-output couples. In this context, a surprising yet often observed property of neural networks is their capacity to provide statistically correct outputs even for unknown inputs. In spite of the ubiquity of this phenomenon known as \textit{generalization} it is still far from being completely understood and the principles and tools of physics can certainly help in this quest \cite{watkin1993statistical,carleo2019machine,zdeborova2020understanding,goldt2020modeling}. In particular, early studies of shallow neural networks have shown that a phase transition to error-free generalization can take place for perfectly learnable rules \cite{gyorgyi1990first,sompolinsky1990learning,seung1992statistical}. This phenomenon has recently received the name of \textit{grokking} as it was re-observed in deep neural networks \cite{power2022grokking} and subsequently analyzed as a phase transition in a shallow network \cite{rubin2023droplets}. Grokking, understood as a phase transition of a network towards a perfect generalization state, therefore emerges as a very general phenomenon in learning, neither limited to algorithmic data \cite{liu2022omnigrok} nor to specific architectures or training processes \cite{kinzel1998phase,vzunkovivc2024grokking,miller2024grokking}. It is however well established that the generalization property of neural networks comes at the cost of their propension to forget older inputs, a phenomenon known as catastrophic forgetting \cite{french1999catastrophic,kirkpatrick2017overcoming}. Far from an implementation detail, this situation results from a fundamental compromise to be found in the stability-sensitivity dilemma \cite{hebb1949organization}. 

In the following, we demonstrate that a local optimization rule (as opposed to a global minimization procedure) can drive the error-free evolution of a network towards a perfect generalization state. Correspondingly, the learning dynamics shows distinctive features of a phase transition. 

With this approach, we strike a new balance in the stability-sensitivity spectrum \cite{french1997pseudo}: lookup tables (LUT) are perfect memories which lack any generalization capability and at the other end deep neural networks can generalize extremely complex problems but may catastrophically forget older data. Instead, our network evolves under a constraint of error free operation (thus ruling out forgetting) and dynamically transfers complexity from the nodes to the network topology, which provides generalization capability. 

Our approach strongly differs from standard artificial neural networks in several ways. First, there is no global cost function such as the training losses of statistical learning. On the contrary, our network dynamics is ruled by a \textit{local} minimization principle, as each node evolves based only on locally available information. Besides the long-questioned biological plausibility of a global minimization procedure (see \textit{eg} \cite{grossberg1987competitive,lillicrap2020backpropagation}), this is a fundamental difference since the dynamics we propose is not variational, as opposed to the usual training by gradient descent. Second, we do not update continuously tunable weights since our network evolves via the addition, merger or deletion of nodes and binary edges. Therefore, we do not fit an existing network to training data. Instead, the topology of our network evolves dynamically upon arrival of new data. Self-organized networks whose topology may evolve (see \textit{eg} \cite{marsland2002self,horzyk2004self,parisi2017lifelong,wiwatcharakoses2020soinn+}) are often considered in the context of continual learning \cite{parisi2019continual}. In this context, the focus is less on grokking than on adaptation to new data while avoiding catastrophic forgetting. Here instead, the network evolution involves first perfectly storing new data by adapting a node (which consist of a LUT) and then transferring the complexity from the individual nodes to the network, the evolution of each node being driven by greedy optimization.

The paper is organized as follows. In section \ref{Modelisation} we describe the nature of our proposed network and the processes which drive its evolution during learning. In \ref{LearningDynamics} we illustrate the resulting dynamics with the example of a network learning to solve the N-bit parity problem towards the final state of perfect generalization. In \ref{Statistics}, we analyze the statistics of the generalization error with respect to the dimension of the problem and discuss it in relation with a phase transition. Finally, in section \ref{Examples} we show that the proposed algorithm can generate learning networks capable of grokking many binary classification tasks, including the primality problem. We discuss the implications of our results in \ref{sec:discussion} and conclude in \ref{Conclusion}.

\section{Modelisation and algorithm}
\label{Modelisation}
In deep learning algorithms, a distinction is made between the training phase, where the network is asked to reproduce the set of data provided as examples (training data samples or TDS), and the generalization phase, where the network is able to correctly predict the result on data it has not yet learned. The training threshold precedes the generalization threshold, but both are obtained by increasing the number of iterations of the same global error minimization procedure \cite{power2022grokking}, wrongly suggesting that the generalization threshold is concomitant with better accuracy in the learning phase.

Here we design a learning network model in which the training phase is simplified to the extreme, as it simply disappears: Each new TDS is automatically, instantaneously and perfectly acquired on the fly. On the other hand, each new learning phase is followed by a network remodeling phase, during which the information learned is judiciously redistributed between the nodes and the information flow restructured by the topology of the connections. All the organization rules are local: local in time, as they act after each new TDS, and local in space, as they involve a single node or pair of nodes. 

Our algorithm calls on multiple interdependent concepts, whose respective roles can only be understood through the input and output of the other concepts. A sequential presentation is not well adapted to this recurrent structure and we have therefore decided to present our algorithm in two complementary, if slightly redundant, stages. First, we give a synthetic description: we briefly introduce the various basic concepts (what the network is made of, the dynamics of information on this network, the network's own dynamics with the notions of splitting and merging, learning a new rule, conflict resolution and cleaning) and we describe the syntactic structure of the algorithm i.e. how these basic concepts relate to each other. The second stage takes up each of the concepts introduced in the first part and describes them in detail. Particular emphasis is placed on the node splitting mechanism and how it underpins the local generalization process at the heart of our algorithm. This second step may be omitted by the impatient reader on first reading.

\begin{figure}[!h]
\includegraphics[width=0.35\textwidth]{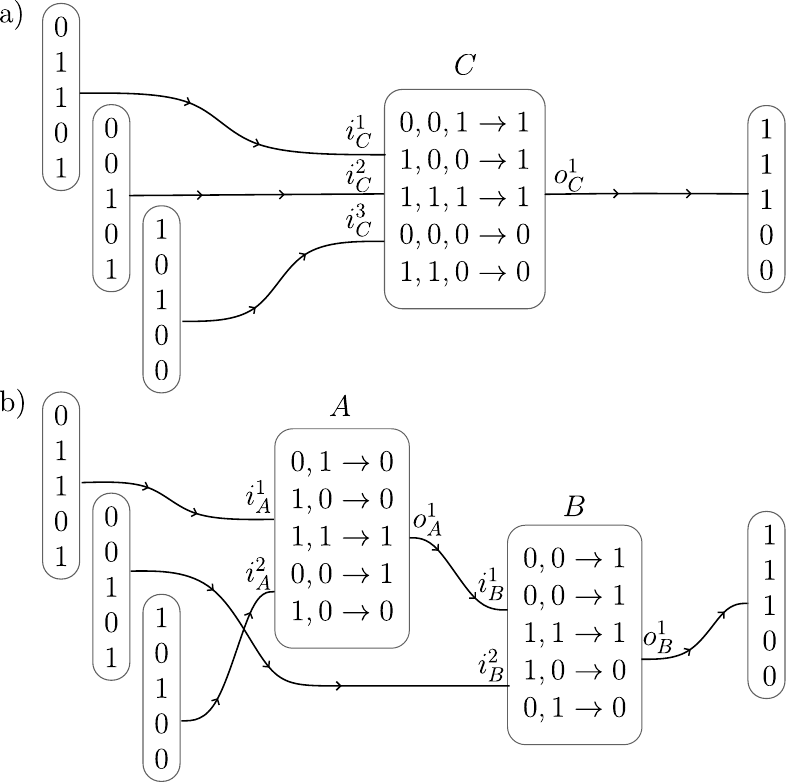}
	\caption{Network before (a) and after (b) topological evolution for the $N-$bits parity problem with $N=3$. There are $4$ interface nodes ($3$ input, $1$ output), $1$ hidden node for (a) and $2$ for (b). There are $5$ TDS, each associated with a line of the various tables involved. The notation $i_{A}^{p}$ is used to designate the $p$ th input connection to $A$, $o_{A}^{q}$ for its $q$ th outgoing connection. As an example of how to read figure (a), we consider the case of the first (resp. $2$nd) training sample: the top left node sends $0$ (resp. $1$) , the middle left node $0$ (resp. $0$) and the bottom left node $1$ (resp. $0$). The central node then receives the code $0,0,1$ (resp. $1,0,0$). According to the node's LUT, the corresponding output code is $1$ (resp. $1$), which is retrieved by the last node. Figure (b) reads the same way, from left to right, layer by layer. It is important to note that networks (a) and (b) are strictly equivalent for all $5$ training samples, but that network (b) is able to provide a correct answer to all the $2^N$ cases, including the three which are not part of the training samples (cf. text).}
\label{fig:network}
\end{figure}

\subsection{Synthetic description}

\subsubsection{What the network is made of}
The network is a directed graph made up of nodes and links. The nodes are arranged on
layers. The $N_{e}$ nodes on the first layer and the single node on the last layer have a special role: The former (left side) have only one outgoing connection, the latter (right side) only one incoming connection. We'll call them interface nodes. The $N$ nodes between these two layers form the hidden nodes. During the network dynamics phase, the number, connections and behavior of the hidden nodes change. Interface nodes, on the other hand, are immutable in the sense that network dynamics preserve the uniqueness of outgoing and incoming connections respectively. We'll call last hidden node and note (LHN), the only hidden node connected to the output interface node.
The similarity with neural systems ends here. The dynamics of a true numerical neuron (and, to a lesser extent, a biological neuron) are quite straightforward. A single action potential is calculated, and the same response is sent to all output links. Action potential and response are linked by a $\mathbb{R}  \longrightarrow \mathbb{R}$ activation function. All the informations associated with the training data are stored in the network in the form of connection weights. In contrast, in our network, the informations associated with the TDS are stored partly in the network in the form of the topology of connections, but also partly in each node ($i$) in the form of a multidimensional response function $\{0,1\}^{nil(i)} \longrightarrow \{0,1\}^{nol(i)}$ specific to each node, where $nil(i)$ denotes its number of incoming links and $nol(i)$ its number of outgoing links.

\subsubsection{ Dynamics on the network}
\label{DynamicsOnNetwork}
Fig.\ref{fig:network} deals with an elementary well known machine learning playground consisting in determining the odd or even
nature of the sum of the $Ne = 3$ components of a binary
vector. The figure shows the situation after $5$ TDS (hence the stack of $5$ horizontal lines). The boxes associated with the input interface nodes correspond to the output codes of these
nodes during each of the $5$ TDS. The box for the output interface node lists the input codes for this node for all $5$ TDS.

Each node in the network behaves like a LUT, matching any input code with a single output code. For example the node $A$ in fig.\ref{fig:network}b has $2$ incoming links $i_{A}^{1}$ and $i_{A}^{2}$ from nodes $(1,3)$ on its left as inputs, and $1$ outgoing link $o_{A}^{1}=i_{B}^{1}$ to nodes $(B)$ on its right ($nol$ can be greater than $1$, but it is not the case here due to the extreme simplicity of the elementary problem under consideration). Any input code of $A$ consists of an ordered list of bits ($0$ or $1$), one bit for each input connection. The $A$ node recognizes the current input code as one of the input codes in its LUT and transmits through its output connections the information bits of the output code corresponding to the identified input code. For example for the $1$st TDS (resp. $2$nd), the current input code is $[0,1]$ (resp. $[1,0]$), it is identified as the learning $1$ (resp. $2$) in the LUT and then $A$ transmits the code $[0]$ (resp. $0$) to $B$. In fig.\ref{fig:network}b, when node $A$ have transmitted its output code, then node $B$ is in possession of the complete input code $[0,0]$ (resp. $[0,0]$) and can then transmit the output code $[1]$ (resp. $[1]$) corresponding to its own LUT. Information thus flows from left to right, and all the nodes form a feed-forward network. 

A TDS for the network consists of a set of output codes for the interface nodes. A learning for a hidden node consists of a pair of input and output codes for its connections. Despite the similarities between the two concepts, we'll use 2 distinct words to emphasize the difference in scale, but above all in data origin. The network receives unchanging data from outside, while the data accessible to a node has been processed by upstream nodes. The tricky part is mapping each network TDS to the multiple hidden nodes learning (going from (a) to (b) in fig.\ref{fig:network}), but later (in \ref{NetorkDynamics}) we'll discuss an algorithm that does this without loss of information. But now, the result is that each hidden node ends up storing information in the form [ learning number, input code, output code] . The length of the node LUT therefore increases by $1$ unit with each new network TDS, and the same pair [entry code, output code] may be repeated several times. As a result, the learning process is straightforward: there are no connection weights to adjust, there is no non-linear thresholding function, there is no global minimization of error. All that is needed is to add a new item to the LUTs such that every new learning is exactly and immediately acquired.

\subsubsection{Network's own dynamics}
\label{NetorkDynamics}
Once the new information has been rapidly acquired, the network topology may change through mechanisms, splitting and merging, that are both local ( i.e. they only involve processes associated with one node or a pair of nodes) and conservative (the training error is exactly zero throughout the processes). 

Splitting and merging are illustrated in fig.\ref{fig:network}. Splitting is the conservative process whereby a hidden node $C$ splits into 2 nodes $A$ and $B$, sharing incoming and outgoing links and possibly creating links between $A$ and $B$. Merging is the opposite conservative process, in which the two hidden nodes $A$ and $B$ merge to form a single node $C$. Both elementary processes can be chained together at will to profoundly modify the network topology, again with no effect on the training error. We discuss later on how these mechanisms can lead to a much more complex topology for more complex problems (see fig.\ref{fig5})

Splitting and merging are the basic antagonistic tools of topological transformations. When and how should they be used? Splitting is associated with data analysis and is responsible for proposing a coherent syntactic structure through inductive reasoning. It enables the network to propose a coherent (but not necessarily correct) interpretation of the local data. Merging occurs when a new TDS to be learned turns out to conflict with the previously coherent interpretation of the data. It allows us to backtrack.

The splitting action replaces the central node $C$ in (a) with $2$ hidden nodes $A$ and $B$ in (b), positioned on $2$ distinct layers (fig.\ref{fig:network}). Local network dynamics are obviously modified, but the reader is strongly encouraged to check that (b) faithfully reproduce  the  $5$ TDS described in (a). Absolutely no training information was lost during the splitting operation. Now it's crucial to note that since the central node $C$ has $3$ incoming links, its LUT will only be complete after $2^3$ independent learnings. Indeed, node $C$ in fig.\ref{fig:network}a doesn't know any of the codes $[0,1,0]$, $[0,1,1]$ and $[1,0,1]$. Not only has it never encountered them before, but it's also unable of coming up with a plausible answer. On the contrary, we can check point by point that nodes $A$ and $B$ in the new topology (fig.\ref{fig:network}b) are really able to propose an answer. We emphasize that the proposed answer could have been wrong. Instead, the network provides here a correct answer for all the $2^3$ possible codes, including the $3$ unknown ones.

Finally, note that splitting can be chained. However, in this example, splitting one or both hidden nodes doesn't add anything.

\subsubsection{New training sample and conflict}
\label{NewLearn}

Consider a network with $N_{e}$ input interface nodes, which has already been taught $m$ TDS. When we measure a non-zero generalization error, we deduce that we must continue to feed the network with new TDS. We therefore supply the left interface nodes with the output codes corresponding to the new training sample to be learned, and observe their evolution according to the dynamics described in \ref{DynamicsOnNetwork}. The input code seen by the nodes during this propagation is referred to as “current”. 

For values of $m$ smaller than $2^{N_{e}}$, it often happens that one of the hidden nodes receives a current input code that is unknown to it, i.e. that doesn't correspond to any of the entries in its LUT. This current input code is then simply learned by integration into the LUT. In the case of LHN, the output code to be learned is the output code of the new network TDS.  In the case of a hidden node other than LHN, an additional connection is created to each outgoing node to inform them of the presence of an unknown code. It transmits a bit $0$ for old codes and $1$ for the current input code, the other output bits remain unchanged in the first case, or are completed by analogy with the nearest known example in the second case. This procedure is then chained, from the first hidden node to receive an unknown code to all its downstream nodes.

For larger values of $m$, it may happen that the new TDS input propagates until the network output, with all current input codes known, but the current output code of LHN does not match the network TDS output to be learned. This causes a conflict. By construction, the conflict is not between the new network TDS and the previous $m$ network TDS already learned, but rather between the new TDS and the coherent interpretation of the structure of the previous $m$ achieved by successive splitting. This is where the 2-nodes merging procedure comes into play. Thanks to merging, we then progressively unravel the splitting procedure, starting with the ascending nodes closest to the conflicting one. The merging procedure continues until the new current entry code, now fully fleshed out, becomes unknown.

\subsubsection{Cleaning} 
\label{Cleaning}
A node may have incoming connections that carry no relevant information. These redundant connections are useless, but they do add to the computational burden. Each node must therefore detect and eliminate them. Also this pruning may lead to the creation of hidden nodes with no outgoing connections. When these nodes are not connected to the interface nodes, they must self-destruct.

\subsubsection{Algorithm}
Learning the network up to the transition to grokking is then a dynamic process involving the various concepts previously introduced in a recurrent loop :
\begin{enumerate}
\item We measure the generalization error $E_{g}$ of a network that has learned $m$ TDS. If this error is zero, nothing more needs to be done. Otherwise, an additional training sample $m \longrightarrow m+1$ must be learned. Any conflicts are resolved as described in \ref{NewLearn}. The number of nodes and layers can be reduced during this phase.
\item Then, as long as this leads to a local improvement in generalization, a splitting procedure is applied to one of the network's hidden nodes taken at random. This phase increases the number of nodes and layers.
\item Clean up the network.
\item Return to point 1.
\end{enumerate}

In section \ref{Examples} we'll give a number of examples of successful applications of this algorithm. The problems addressed and their algorithmic complexity are very varied.

\subsection{Detailed Description}

\subsubsection{2-nodes merging}
\begin{figure}[!h]
\resizebox{0.40\textwidth}{!}{
\includegraphics[]{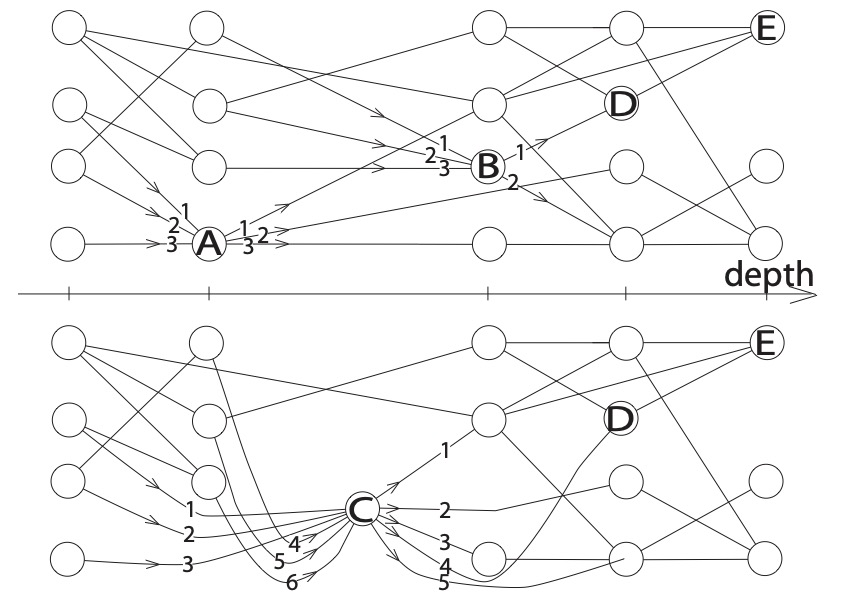}
}
\caption{ Network configuration before (top) and after (bottom) the merging $(A,B) \longrightarrow C $. For node $A$, the left-hand $p$ and right-hand $q$ numbers overlapping the connections, label the incoming and outgoing links noted $i_{A}^{p}$ and respectively $o_{A}^{q}$ in the text. For the sake of clarity, we have used here a simplified notation.
It is also important to note that all the pairs of nodes can not be grouped. This is the case, for example, of nodes $B$ and $E$ in the top figure.}
\label{fig2}
\end{figure}

The mechanism for merging two nodes is fairly intuitive and easy to understand. It is illustrated in fig.\ref{fig2} for the topology and tab.\ref{tab1} for the LUTs. In fig.\ref{fig2} top, before merging, node A has input connections $i_{A}^{1}$, $i_{A}^{2}$ and $i_{A}^{3}$ and output connections $o_{A}^{1}$, $o_{A}^{2}$ and $o_{A}^{3}$. $B$ has inputs connections $i_{B}^{p}$, $p \in [1,3]$ and outputs connections $o_{B}^{q}$, $q \in [1,2]$. After merging (fig.\ref{fig2} bottom), the resulting node $(A,B) \longrightarrow C$ has inputs codes constructed simply by concatenating the input codes of the initial nodes $A$ and $B$, as shown in tab.\ref{tab1}. The same mechanism applies to output codes.

Two remarks are now in order. First in the example in fig.\ref{fig2} and tab.\ref{tab1}, nodes $A$ and $B$ are not directly connected. But the merging procedure can easily be generalized to cases where $A$ and $B$ are connected, as in the example in fig.\ref{fig1} and tab.\ref{tab2} (provided you start from fig.\ref{fig1}b and go to fig.\ref{fig1}a). In this case, the internal links between $A$ and $B$ disappear, and the information bits associated with them are not taken into account when the codes are concatenated. Second, it is important to realize that not all node pairs can be merged. This is the case, for example, with nodes $B$ and $E$ in fig.\ref{fig2}. In this configuration, node $D$ is both an output of $B$ and an input of $E$, and therefore should appear as both an input and an output of their merger. This would be possible only in a recurrent network and we choose to restrict our analysis to a feedforward network.

\begin{table}[h!]
\centering
\begin{tabular}{|c|c|c|}
 \hline
Learning: A&  Input codes  &  Output codes  \\
 \hline
1 &
\begin{tabular}{ccccc}
1 & 1 & 1 
\end{tabular}
&
\begin{tabular}{ccc}
0 & 0  & 1
\end{tabular}
\\
\hline
2 &
\begin{tabular}{ccccc}
0 & 0 & 1 
\end{tabular}
&
\begin{tabular}{ccc}
0 & 1  & 0
\end{tabular}
\\
\hline
3 &
\begin{tabular}{ccccc}
1 & 0 & 0 
\end{tabular}
&
\begin{tabular}{ccc}
0 & 0  & 0
\end{tabular}
\\
\hline
4 &
\begin{tabular}{ccccc}
0 & 1 & 0 
\end{tabular}
&
\begin{tabular}{ccc}
1 & 0  & 0
\end{tabular}
\\
\hline
5 &
\begin{tabular}{ccccc}
0 & 0 & 0 
\end{tabular}
&
\begin{tabular}{ccc}
1 & 1  & 0
\end{tabular}
\\
\hline
Learning: B &  Input codes  &  Output codes  \\
 \hline
1 &
\begin{tabular}{ccccc}
1 & 1  & 1 
\end{tabular}
&
\begin{tabular}{ccc}
1 &  0
\end{tabular}
\\
\hline
2 &
\begin{tabular}{ccccc}
0 & 0 &  0
\end{tabular}
&
\begin{tabular}{ccc}
1   & 0
\end{tabular}
\\
\hline
3 &
\begin{tabular}{ccccc}
0 & 1 &  0 
\end{tabular}
&
\begin{tabular}{ccc}
0 &  1
\end{tabular}
\\
\hline
4 &
\begin{tabular}{ccccc}
1 & 0 & 0 
\end{tabular}
&
\begin{tabular}{ccc}
1  & 0
\end{tabular}
\\
\hline
5 &
\begin{tabular}{ccccc}
0 & 1 &  0 
\end{tabular}
&
\begin{tabular}{ccc}
0 &  1
\end{tabular}
\\
\hline
Learning: C &  Input codes  &  Output codes  \\
 \hline
1 &
\begin{tabular}{ccccccccc}
1 & 1  & 1  &1 &1 &1  
\end{tabular}
&
\begin{tabular}{ccccc}
0 &  0 &  1 & 1 & 0
\end{tabular}
\\
\hline
2 &
\begin{tabular}{ccccccccc}
0 & 0  & 1  &0 &0 &0  
\end{tabular}
&
\begin{tabular}{ccccc}
0 &  1 &  0 & 1 & 0
\end{tabular}
\\
\hline
3 &
\begin{tabular}{ccccccccc}
1 & 0 & 0  &0 &1 &0  
\end{tabular}
&
\begin{tabular}{ccccc}
0 &  0 &  0 & 0 & 1
\end{tabular}
\\
\hline
4 &
\begin{tabular}{ccccccccc}
0 & 1  & 0  &1 &0 &0  
\end{tabular}
&
\begin{tabular}{ccccc}
1 &  0 &  0 & 1 & 0
\end{tabular}
\\
\hline
5 &
\begin{tabular}{ccccccccc}
0 & 0  & 0  &0 &1 &0  
\end{tabular}
&
\begin{tabular}{ccccc}
1 &  1 &  0 & 0 & 1
\end{tabular}
\\
\hline
\end{tabular}
\caption{Input and output codes associated with nodes $A$, $B$ and $C$ in fig.\ref{fig2}.}
\label{tab1}
\end{table}

\begin{table}[h!]
\centering
\begin{tabular}{|c|c|c|}
 \hline
Learning: C &  Input codes  &  Output codes  \\
 \hline
1 &
\begin{tabular}{cccccc}
1 & 0  & 0 & 1 & 0 &0 
\end{tabular}
&
\begin{tabular}{cccc}
1 &  1 &  0 & 0 
\end{tabular}
\\
\hline
2 &
\begin{tabular}{cccccc}
0 & 1  & 1 & 0 & 1 & 0  
\end{tabular}
&
\begin{tabular}{cccc}
1 &  0 &  0 & 1
\end{tabular}
\\
\hline
3 &
\begin{tabular}{cccccc}
0 & 1 & 0 & 1 & 1 & 1 
\end{tabular}
&
\begin{tabular}{cccc}
0 &  0 &  0 & 1 
\end{tabular}
\\
\hline
4 &
\begin{tabular}{cccccc}
1 & 1  & 1 & 1 & 0 & 0 
\end{tabular}
&
\begin{tabular}{cccc}
1 &  1 & 1 & 0 
\end{tabular}
\\
\hline
5 &
\begin{tabular}{cccccc}
0 & 0  & 1 & 0 & 1 & 0 
\end{tabular}
&
\begin{tabular}{cccc}
0 &  1 &  0 & 0 
\end{tabular}
\\
\hline
Learning: A&  Input codes  &  Output codes  \\
 \hline
1 &
\begin{tabular}{cccc}
1 & 0 & 0 & 1 
\end{tabular}
&
\begin{tabular}{cccc}
1 &1 & 0  & \fcolorbox{SkyBlue}{SkyBlue}{0}
\end{tabular}
\\
\hline
2 &
\begin{tabular}{cccc}
0 & 1 & 1 & 0 
\end{tabular}
&
\begin{tabular}{cccc}
1 & 0 & 1  & \fcolorbox{SkyBlue}{SkyBlue}{0}
\end{tabular}
\\
\hline
3 &
\begin{tabular}{cccc}
0 & 1 & 0 & 1 
\end{tabular}
&
\begin{tabular}{cccc}
0 & 0 & 1  & \fcolorbox{SkyBlue}{SkyBlue}{0}
\end{tabular}
\\
\hline
4 &
\begin{tabular}{cccc}
1 & 1 & 1 & 1 
\end{tabular}
&
\begin{tabular}{cccc}
1& 1 & 0  & \fcolorbox{SkyBlue}{SkyBlue}{1}
\end{tabular}
\\
\hline
5 &
\begin{tabular}{cccc}
0 & 0 & 1 & 0 
\end{tabular}
&
\begin{tabular}{cccc}
0& 1 & 0  & \fcolorbox{SkyBlue}{SkyBlue}{0}
\end{tabular}
\\
\hline
Learning: B &  Input codes  &  Output codes  \\
 \hline
1 &
\begin{tabular}{ccc}
\fcolorbox{SkyBlue}{SkyBlue}{0} & 0  & 0 
\end{tabular}
&
\begin{tabular}{c}
0
\end{tabular}
\\
\hline
2 &
\begin{tabular}{ccc}
\fcolorbox{SkyBlue}{SkyBlue}{0} & 1 &  0 
\end{tabular}
&
\begin{tabular}{c}
0
\end{tabular}
\\
\hline
3 &
\begin{tabular}{ccc}
\fcolorbox{SkyBlue}{SkyBlue}{0} & 1 &  1 
\end{tabular}
&
\begin{tabular}{c}
0
\end{tabular}
\\
\hline
4 &
\begin{tabular}{ccc}
\fcolorbox{SkyBlue}{SkyBlue}{1} & 0 & 0 
\end{tabular}
&
\begin{tabular}{c}
1
\end{tabular}
\\
\hline
5 &
\begin{tabular}{ccc}
\fcolorbox{SkyBlue}{SkyBlue}{0} & 1 &  0 
\end{tabular}
&
\begin{tabular}{c}
0
\end{tabular}
\\
\hline
\end{tabular}
\caption{Input and output codes associated with nodes $A$, $B$ and $C$ in fig.\ref{fig1}. The bits on a blue background are associated with the additional connection between $A$ and $B$. They disappear in the $C$ merger codes.}
\label{tab2}
\end{table}

\subsubsection{2-nodes splitting}

The 2-nodes splitting procedure is illustrated in fig.\ref{fig1} for the topology and tab.\ref{tab2} for the LUTs: it's an operation that allows you to switch from configuration (a) with a single node (C) to configuration (b) with 2 nodes (A and B), while maintaining the training error at exactly zero (tab.\ref{tab2}). The transformation leaves the inputs and outputs (connections and nodes) invariant, such that it can be performed locally without any major changes to the surrounding nodes. During the procedure, every incoming connection to $C$ becomes an incoming connection to $A$ or (exclusive) $B$. The same applies to outgoing connections. Possible additional connections from $A$ to $B$ may appear, giving nodes $A$ and $B$ different roles. Finally, the operation can be completely cancelled by applying the 2-nodes merging procedure.

The advantage of 2-nodes splitting lies in the fact that there are multiple ways of dividing the incoming (resp. outgoing) connections of $C$ into 2 disjoint sets, and that we can therefore make a "judicious" choice among the available configurations $(A,B)$ without incurring any change in the training error.

\begin{widetext}
\begin{figure*}[t]
\resizebox{1.00\textwidth}{!}{
\includegraphics[]{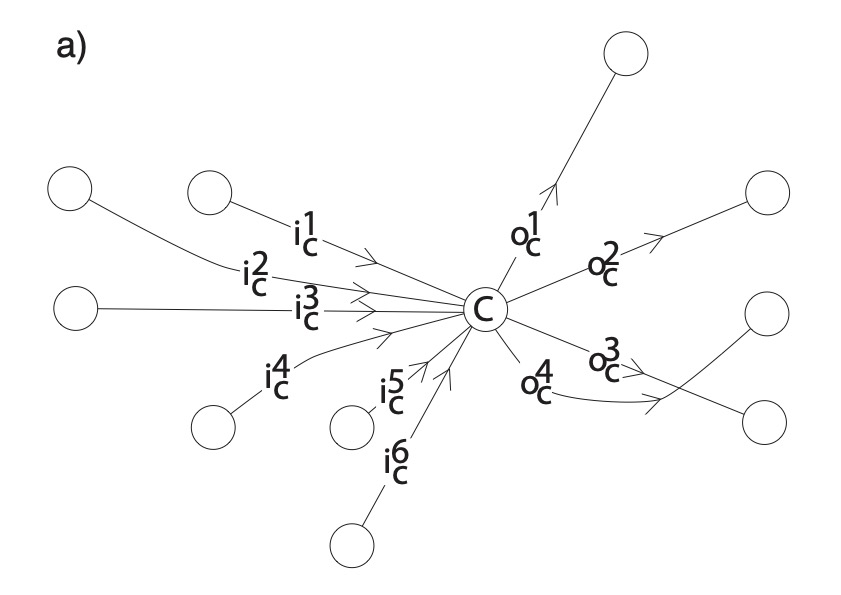}
\includegraphics[]{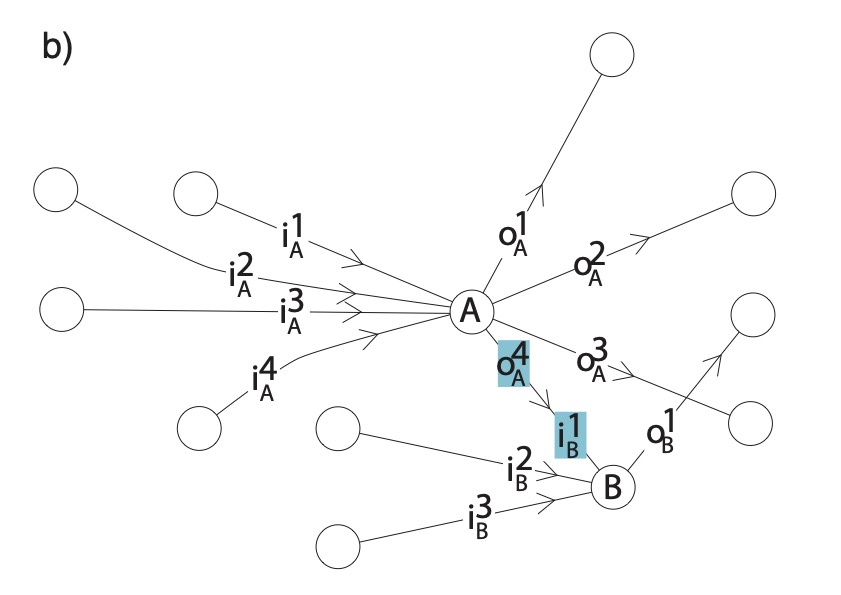}
}
\caption{ Schematic representation of the splitting procedure $C \longrightarrow (A,B)$. The diagram (a) shows the initial node $C$ surrounded by its 6 inputs ($i_{C}^{p}, p \in [1,6]$) and 4 outputs ($o_{C}^{q}, q \in [1,4]$) connections. Plot (b) display the configuration after the splitting procedure. Note the extra link between $A$ and $B$ on the blue background, labeled $i_{B}^{1}$ or $o_{A}^{4}$   depending on whether it's considered as an ingoing or outcoming connection. This link carries the extra bits shown on the blue background in tab.\ref{tab2}}.
\label{fig1}
\end{figure*}
\begin{figure*}
\resizebox{1.00\textwidth}{!}{
\includegraphics[]{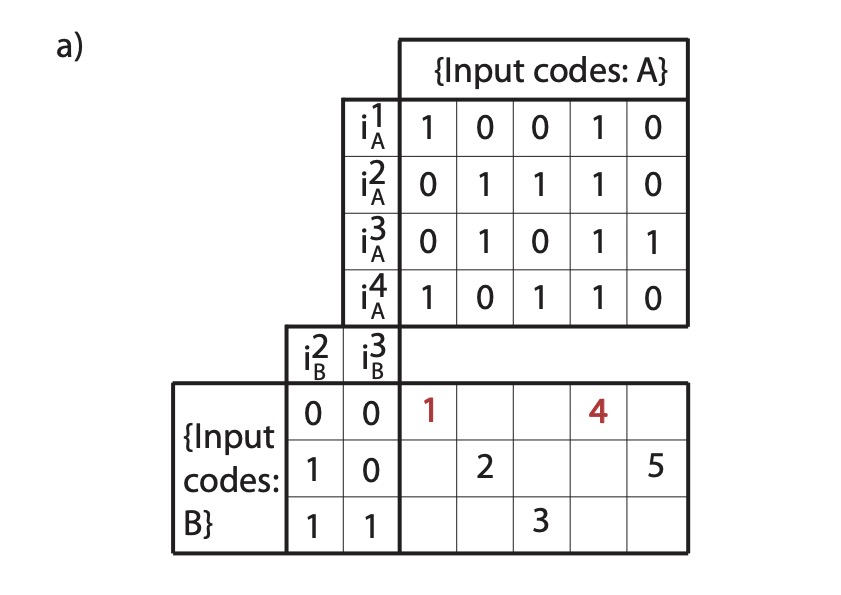}
\includegraphics[]{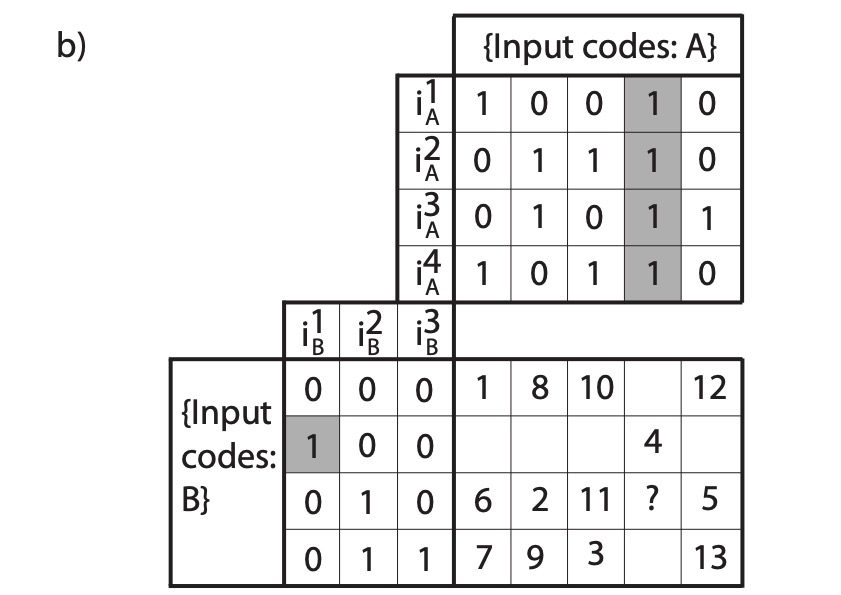}
}
\caption{ Table of input codes for nodes $A$ and $B$ after $m=5$ training data learnings. (a): Numbers from $1$ to $5$ correspond to the projection on $A$ and $B$ of the $C$ input codes for the $5$ training data. Because of $1$ and $4$ (in red on (a)), the LUT of $B$ does not define an application. We correct this by introducing an additional link $o_{A}^{4}=i_{B}^{1}$ (elements involved on a gray background in (b)). Numbers from $6$ to $13$ then correspond to consistent (although perhaps incorrect) responses from the pair ($A,B$) for input codes unknown to $C$.}
\label{fig3}
\end{figure*}
\end{widetext}

Let's start by identifying the possible configurations. Note that in the configuration fig.\ref{fig1}a, the output codes of node A do not depend at all on the responses of node B. This means that A's LUT is an application in the mathematical sense of the term, i.e. each input code of $A$ has one and only one output image. This imposes a very strong constraint on the choice of input and output connections for $A$. For example, let's suppose we've selected for $A$ the input connection  $i_{A}^{1}$ and $i_{A}^{2}$, and the output connection $o_{A}^{1}$. The LUT for $A$ would then be given by tab.\ref{tab3}. Clearly, the same input code $[0,1]$ would then have been associated with two distinct output codes, depending on whether we had considered the $2$ or $3$ learnings. As this configuration was not an application, it had to be rejected.

\begin{table}[h!]
\centering
\begin{tabular}{|c|c|c|}
\hline
Learning: A & Input codes & Output codes
\\
\hline
1 & \begin{tabular}{cc} 1 & 0 \end{tabular} & 1
\\
\hline
2 & \begin{tabular}{cc} 0 & 1 \end{tabular} & 1
\\
\hline
3 & \begin{tabular}{cc} 0 & 1 \end{tabular} & 0
\\
\hline
4 & \begin{tabular}{cc} 1 & 1 \end{tabular} & 1
\\
\hline
5 & \begin{tabular}{cc} 0 & 0 \end{tabular} & 0
\\
\hline
\end{tabular}
	\caption{Example of an impossible splitting choice. LUT of $A$ for the configuration in fig.\ref{fig1}, when $A$'entries are limited to $i_{A}^{1}$ and $i_{A}^{2}$, while $A$'s exit to $o_{A}^{1}$.}
\label{tab3}
\end{table}

Once we've identified a set of valid inputs and outputs connections for $A$, we can naturally consider a first definition of inputs and outputs for $B$: those for $C$ minus those for $A$. At this point, the various LUTs are then described by tab.\ref{tab2}, deprived of the $2$ columns on a blue background and located respectively in the “Learning: $A$, Output codes, last column” and “Learning: $B$, input codes, first column” areas. But it turns out that the LUT of $B$ thus constructed is not an application: the same input code [0,0] appears $2$ times, for learning $1$ and $4$, but associated with distinct output codes. As the LUT of $C$ is an application, it should be enough to retrieve some of the information stored in $A$ to be able to distinguish the 2 input codes of $B$. This is the role assigned to the additional connection between $A$ and $B$ (blue background columns in tab.\ref{tab2}). It transmits $0$ for all learnings except $4$, for which it transmits $1$, thus lifting the degeneracy.

We've just seen how to perform the 2-nodes splitting procedure. Now we come to the fundamental question: what's in it for us? As the gain is not to be found in the training error, since splitting does not modify it in any way, we focus on the generalization error. To do this, we'll compare the responses of the pair $(A,B)$ and those of $C$ to the presentation of an unknown current input code. We begin by constructing the table in fig.\ref{fig3}(a) of cross-referenced input codes for $A$ and $B$. Until now, we've been working with lists of length $m$ of input codes, where $m$ is the number of training data. Here, we use the notation $\{$input codes of $A$$\}$ to refer to the set of input codes of $A$ (idem for $B$). The difference is that in a set, unlike a list, an element is not allowed to repeat itself. The $m=5$ input codes of $C$ break down into the input codes of $A$ and $B$ as described in tab.\ref{tab2} and correspond to the intersections in fig.\ref{fig3}a marked by the digits $1$ to $m$. It is important to note that the $[0,0]$ and $[1,0]$ entry codes of $B$ appear 2 times, respectively for learnings $1$, $4$ and $2$, $5$. For the $2$ and $5$ learnings, the $B$ output codes associated with $[1,0]$ are identical (tab.\ref{tab2}) and therefore present no difficulties. On the other hand, learnings $1$ and $4$ are problematic because the output codes associated with $[0,0]$ are not identical, and the LUT of $B$ defined by  fig.\ref{fig3}a is not an application. The solution, as described in tab.\ref{tab2}, is to create an additional link from $A$ to $B$ to remove the degeneracy. In  fig.\ref{fig3}b, this corresponds to the column $i_{B}^{1}$. Now consider the current code $[1,0,0,1,1,0]$, unknown to $C$. Decomposed on the pair ($A,B$), it corresponds to the input code $[1,0,0,1]$ of $A$ and $[i_{B}^{2},i_{B}^{3}]=[1,0]$ identified by number $6$ in fig.\ref{fig3}b. The input code $[1,0,0,1]$ of $A$ is associated with the output code $[1,1,0,0]$ (tab.\ref{tab2}). Since $o_{A}^{4}=0=i_{B}^{1}$, we know how to give a value to the output code of $B$ associated with the input code $[0,1,0]$, the one associated with $m=2$ (i.e. $[0]$).  On the other hand, consider the current code $[1,1,1,1,1,0]$ unknown to $C$. Decomposed on the pair ($A,B$), it corresponds to the input code $[1,1,1,1]$ of $A$ and $[i_{B}^{2},i_{B}^{3}]=[1,0]$ marked with “?” in table (b). For this input code of $A$, the answer is $[1,1,0,1]$ (tab.\ref{tab2}) and so we have $o_{A}^{4}=1=i_{B}^{1}$. But the current code $[1,1,0]$ thus constructed is not an input code of $B$. So we don't know how to assign it a value.

We've just shown that the pair $(A, B)$ is sometimes able to provide coherent answers, whereas the initial node C is unable to do so. These answers are not necessarily accurate in the sense that a future network TDS may disavow them, but they have the merit of existing and proposing a coherent generalization.

In fact, in the particular case of fig.\ref{fig1}, the pair $(A,B)$ can provide a consistent answer in $13$ input cases, $5$ of which are exact (fig.\ref{fig3}). As this number will play an important role in our algorithm, we'll call it the number of consistent responses of the $(A,B)$ pair and will denote it NCR$(A,B)=13$ (note that NCR$(C)=5$). For the same initial node $C$, NCR$(A,B)$  depends on the inputs and outputs repartition between $A$ and $B$. With the idea of maximizing generalization, the algorithm tests several possible repartitions between $A$ and $B$ and selects the one that gives the highest value of NCR$(A,B)$. As the number of possible repartitions grows extremely quickly with the number of connections, we won't perform an exhaustive exploration, but rather a random sampling.

Finally 2-nodes splitting operations can be chained together in a cascade process:
\begin{equation}
\begin{array}{lll}
A_{1} & \longrightarrow &(A_{11},A_{12})
\cr \cr
A_{11} & \longrightarrow &(A_{111},A_{112})
\cr
A_{12} & \longrightarrow &(A_{121},A_{122})
\cr \cr
A_{111} & \longrightarrow &(A_{1111},A_{1112})
\cr
....
\end{array}
\end{equation}
which goes on as long as NCR$(A,B)>$NCR$(C)$, i.e. as long as splitting improves generalization.

\begin{figure}[!h]
\resizebox{0.40\textwidth}{!}{
\includegraphics[]{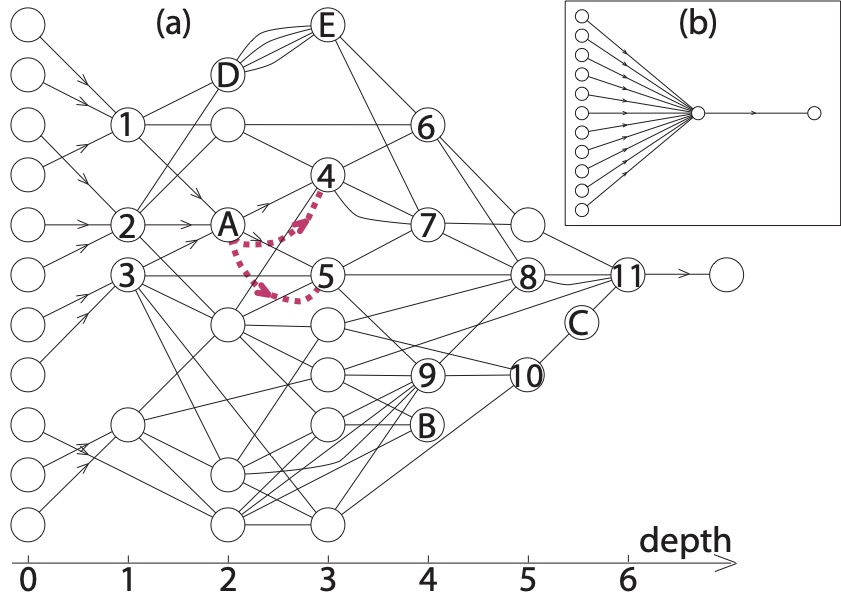}
}
\caption{ A typical network configuration obtained after network topology evolution (a). The insert (b) corresponds to the initial configuration. Node $A$: The dotted red arrows represent the additional connections created by node $A$ on receiving an unknown CIC. Nodes $B$, $C$ and $D$: examples of configurations to be removed during the cleaning operation. }
\label{fig5}
\end{figure}

 \subsubsection{New learning and unknown codes} \label{NewLearningAndUnknownCodes}
 Let's suppose the network is in the configuration shown in fig.\ref{fig5} and has learned a number $m=5$ of TDS. We're now trying to teach it a new TDS. The nodes of the first layer grab the bits of the new input code and pass them on to the nodes of the second layer. Each node of the second layer then receives an input code, its current input code, which it may or may not already know. For the sake of argument, let's assume that all nodes in the second layer know their current input code. We then move on to the third layer. In this layer, let's assume that only node A receives a current input code (CIC($A$)) that it doesn't know. The situation is summarized in tab.\ref{tab7}.

A little nomenclature is useful. We have already introduced $noc(i)$ and $nic(i)$ the number of outgoing and incoming connections of node $i$.
Let $non(i)$ and $nin(i)$ be the number of outgoing and incoming nodes for node $i$. We have $non(i) \le noc(i)$ and $nin(i) \le nic(i)$.

In order to deals with the CIC(A), node $A$ and its outputs are then going to perform 3 operations:
\begin{enumerate}
\item Node $A$ will add an extra connection for each of its outgoing nodes and inform them that they need to increment their $nic$ by one. In our example (fig.\ref{fig5}a):
\begin{equation}
\begin{array}{l}
\left\{
\begin{array}{l}
non(A) \longrightarrow non(A)
\cr
noc(A) \longrightarrow noc(A)+non(A)
\end{array}
\right.
\cr \cr
\left\{
\begin{array}{l}
nin(4) \longrightarrow nin(4)
\cr
nic(4) \longrightarrow nic(4)+1
\end{array}
\right.
\cr \cr
\left\{
\begin{array}{l}
nin(5) \longrightarrow nin(5)
\cr
nic(5) \longrightarrow nic(5)+1
\end{array}
\right.
\end{array}
\end{equation}
\item  Node $A$ will modify its LUT: all its old output codes are completed by a length $non(A)$ list of bits $0$. A's output nodes modify their old input codes by adding 0 (tab.\ref{tab7}).
\item Node A constructs a new current output code (COC(A)): the first part of COC(A) is an approximation obtained by i) identifying the known input code of A closest to CIC(A) ii) then matching its image through the LUT. The second part is a length $non(A)$ list of bit $1$ (tab.\ref{tab7}).
\end{enumerate}
Tab.\ref{tab7} summarize the changes made to the LUTs, while the dotted red lines in fig.\ref{fig5} summarize the changes made to the network. We can check that the transformations made leave the old LUTs invariant, while taking into account the new information.
 \begin{table}[h!]
\centering
\begin{tabular}{|c|c|c|}
\hline
Learning: A & Input codes & Output codes
\\
\hline
1 & \begin{tabular}{ccc} 1 & 0 & 0 \end{tabular} & \begin{tabular}{cccc} 1 & 0 & \fcolorbox{gray!30}{gray!30}{0} & \fcolorbox{gray!50}{gray!50}{0}\end{tabular}
\\
\hline
2 & \begin{tabular}{ccc} 0 & 1 & 1 \end{tabular} & \begin{tabular}{cccc} 0 & 1 & \fcolorbox{gray!30}{gray!30}{0} & \fcolorbox{gray!50}{gray!50}{0} \end{tabular}
\\
\hline
3 & \begin{tabular}{ccc} 0 & 1 & 0 \end{tabular} & \begin{tabular}{cccc} 1 & 1 & \fcolorbox{gray!30}{gray!30}{0} & \fcolorbox{gray!50}{gray!50}{0} \end{tabular}
\\
\hline
4 & \begin{tabular}{ccc} 1 & 1 & 1 \end{tabular} & \begin{tabular}{cccc} 0 & 0 & \fcolorbox{gray!30}{gray!30}{0} & \fcolorbox{gray!50}{gray!50}{0} \end{tabular}
\\
\hline
5 & \begin{tabular}{ccc} 0 & 0 & 1 \end{tabular} & \begin{tabular}{cccc} 1 & 1 & \fcolorbox{gray!30}{gray!30}{0} & \fcolorbox{gray!50}{gray!50}{0} \end{tabular}
\\
\hline
Current & Current input code  & \begin{tabular}{cccc} . & . & \fcolorbox{gray!30}{gray!30}{1} & \fcolorbox{gray!50}{gray!50}{1} \end{tabular}
\\
\hline
Learning: 4 & Input codes & Output codes
\\
\hline
1 & \begin{tabular}{cccc} . & 1 & . & \fcolorbox{gray!30}{gray!30}{0}\end{tabular} & \begin{tabular}{ccc} . & . &. \end{tabular}
\\
\hline
2 & \begin{tabular}{cccc} . & 0 & . & \fcolorbox{gray!30}{gray!30}{0}\end{tabular} & \begin{tabular}{ccc}. & . &.  \end{tabular}
\\
\hline
3 & \begin{tabular}{cccc} . & 1 & . & \fcolorbox{gray!30}{gray!30}{0}\end{tabular} & \begin{tabular}{ccc} . & . &. \end{tabular}
\\
\hline
4 & \begin{tabular}{cccc} . & 0 & . & \fcolorbox{gray!30}{gray!30}{0}\end{tabular} & \begin{tabular}{ccc} . & . &.  \end{tabular}
\\
\hline
5 & \begin{tabular}{cccc} . & 1 & . & \fcolorbox{gray!30}{gray!30}{0}\end{tabular} & \begin{tabular}{ccc} . & . &.  \end{tabular}
\\
\hline
Current input code & \begin{tabular}{cccc} . & . & . & \fcolorbox{gray!30}{gray!30}{1} \end{tabular} & \begin{tabular}{ccc} ? & ? & ?  \end{tabular}
\\
\hline
Learning: 5 & Input codes & Output codes
\\
\hline
1 & \begin{tabular}{cccc} . & 0 & . & \fcolorbox{gray!50}{gray!50}{0}\end{tabular} & \begin{tabular}{ccc} . & . &. \end{tabular}
\\
\hline
2 & \begin{tabular}{cccc} . & 1 & . & \fcolorbox{gray!50}{gray!50}{0}\end{tabular} & \begin{tabular}{ccc}. & . &.  \end{tabular}
\\
\hline
3 & \begin{tabular}{cccc} . & 1 & . & \fcolorbox{gray!50}{gray!50}{0}\end{tabular} & \begin{tabular}{ccc} . & . &. \end{tabular}
\\
\hline
4 & \begin{tabular}{cccc} . & 0 & . & \fcolorbox{gray!50}{gray!50}{0}\end{tabular} & \begin{tabular}{ccc} . & . &.  \end{tabular}
\\
\hline
5 & \begin{tabular}{cccc} . & 1 & . & \fcolorbox{gray!50}{gray!50}{0}\end{tabular} & \begin{tabular}{ccc} . & . &.  \end{tabular}
\\
\hline
Current input code & \begin{tabular}{cccc} . & . & . & \fcolorbox{gray!50}{gray!50}{1} \end{tabular} & \begin{tabular}{ccc} ? & ? & ?  \end{tabular}
\\
\hline
\end{tabular}
\caption{LUTs of nodes $A$,$4$ and $5$ before and after adding the additional links described by dotted red lines in fig.\ref{fig5}. The corresponding additional bits are on gray background, light for $4$ and dark for $5$. Dots (.\@) denote known values whose value does not need to be specified to understand the presentation. "?" stand for unknown ones.
}
\label{tab7}
\end{table}

Now it's the turn of the nodes in the fourth layer. By construction, nodes 4 and 5 of this layer receive a CIC that is unknown to them (uncertainty transmitted to them by A). We then repeat on these nodes the treatment just applied to A. We then pass from one layer to the next, propagating the lack of knowledge. If we reach LHN with an unknown CIC, then we just have to assign it the output code required by the current network learning operation.

 Note that a node that has already been informed by one of its inputs that it is about to receive an unknown CIC, does not need to be informed again by another of its inputs. Hence node 7 on the fifth layer doesn't need to be informed by both 4 and 5 that its CIC is unknown.
 
\subsubsection{Conflicts} \label{Conflits}
During the learning phase, and especially when the number of TDS starts to grow, it can happen that the LHN receives a CIC that it identifies as known, without the COC associated with it being equal to what the network should be learning. 
This is a conflict situation which clearly indicates that the up to now generalization choices made during the 2-nodes splittings and the consequent network topology are no longer compatible with the last TDS learned.
This conflict is resolved by a cascade of 2-nodes merging propagating from the last to the first layer.
The idea is to make as few changes as possible and therefore focus first on the latest generalizations.
We start by merging LHN with one of the nodes of the previous layers. Then the current learning procedure is repeated until the CIC of the last node is computed. Either the conflict has disappeared, and the CICs and COCs can be saved in the LUT of each network node, or it hasn't, and the last node's merging procedure is repeated. Since the application of the 2-nodes merging procedure, repeated a sufficient number of times, transforms the topology of fig.\ref{fig5}a into that of its insert fig.\ref{fig5}b, and since in the latter configuration there is no possible conflict, since  a new TDS process for the network boils down to simply increasing the LUT of the central node, we can be sure that the conflict resolution procedure described above converges.

\subsubsection{Cleaning}
Learning, even when limited to a small number of input interface nodes ($N_{e}$), is a computationally intensive process. This is particularly true of the current version of the algorithm, which is not parallelized. We therefore take great care to remove all local configurations which are clearly irrelevant and unnecessarily burden calculations. In addition to deleting redundant connections and hidden nodes (already discussed in \ref{Cleaning}), we can also delete gateway nodes. In fig.\ref{fig5}a, node $C$ only acts as a gateway between nodes $10$ and $11$, and node $D$ between $1$, $2$ and $E$. They can be replaced by direct connections between $10$ and $11$ (respectively between $1$, $2$ and $E$) and an adequate redefinition of LUTs through composition.

\subsubsection{Generalisation test and no-fail mode}
Recall that a network TDS is characterized by
the output codes of the interface nodes of the first layer and the corresponding input codes of the last node.
Let $m$ be the number of independent TDS the network has already learned. If $N_{e}$ denotes the number of nodes in the first layer, then the total number of possible network input codes is $2^{N_{e}} \ge m$. The generalization error is defined as the average error the network makes when responding to any of the $2^{N_{e}}$ possible input codes.

New learning, unknown codes, conflicts and 2-nodes mergings and splittings, as described above, have been used to impress training information in the network. But when it comes to measure the generalization error, these learning mechanisms are no longer relevant.  Faced with an unknown input code, the aim is no longer to learn it, but to always provide an answer, possibly approximate, to the outgoing nodes.

The no-fail mode corresponds to the node operating behavior during the generalization error measurement phase. A COC is injected at the output of each interface node in the first layer, and the information is propagated from layer to layer. In the presence of a known CIC, the answer is simply given by the LUT. On the other hand, in the presence of an unknown CIC, the node compares it with each entries in its LUT, selects the known input code closest to the CIC (Euclidean distance) and transmits the corresponding output code as a COC. Finally the network response in no-fail mode is then the CIC of the last-layer node. The error is $0$ if the CIC is indeed the one expected, $1$ otherwise. We emphasize that by construction, no-fail mode provides an exact response for any network input codes that have already been learned.

When $m=2^{N_{e}}$, the generalisation error is trivially vanishing. But what we're really interested in, and what we'll call the “grokking transition”, is getting the generalization error to cancel out for values of $r={{m}\over{2^{N_{e}}}}$ well below $1$.

\section{Learning dynamics}
\label{LearningDynamics}
\begin{figure}[!h]
\resizebox{0.40\textwidth}{!}{
\includegraphics[]{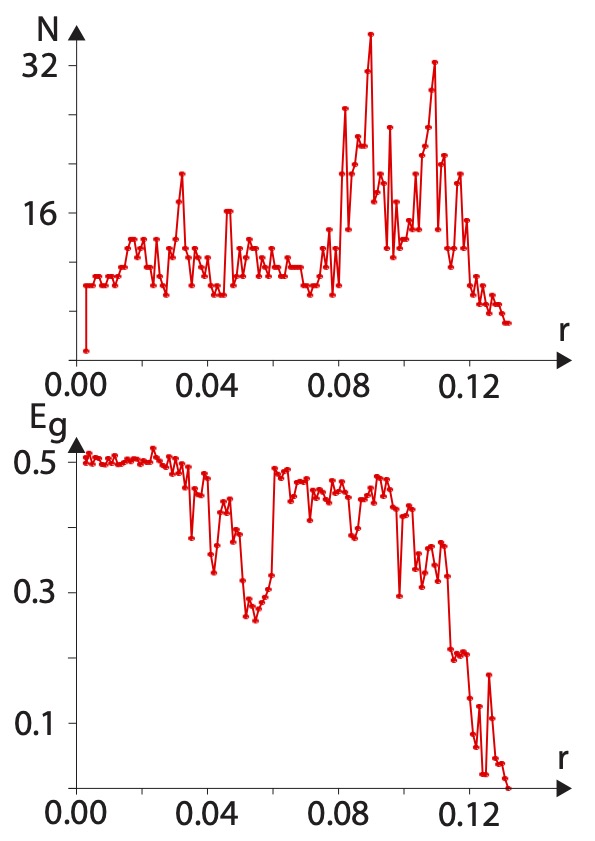}
}
\caption{Dynamics of the algorithm in the case of the parity problem with $N_{e}=10$. The figures show the evolution of the hidden nodes number $N$ (top) and the generalisation error $E_{g}$ (bottom) versus the relative size of the training set $r$. The dots are the numerical measurements while the solid line connecting them is merely a visual aid.}
\label{plan19juillet01}
\end{figure}
We investigate the dynamics of the algorithm, taking as an example the special case of the parity problem: for any input code consisting of $N_{e}$ bits 0 or 1, the network is expected to return $0$ if the number of input bits equal to 1 is even, and 0 otherwise. 
\label{sec:dynamics}

There are $2^{N_{e}}$ possible network entry codes and therefore ${(2^{N_{e}})!}\over{(2^{N_{e}}-m)!}$ ways of choosing $m$ codes from $2^{N_{e}}$, the order of choice being relevant. Such an ordered list of $m$ input codes is called a training set. Fig.\ref{plan19juillet01} is associated with a single, randomly generated training set. The top figure shows the evolution of the number of hidden nodes as a function of $r={{m}\over{2^N_{e}}}$, while the bottom shows the evolution of the generalization error $E_{g}(r)$.

The first thing to notice is that the algorithm really works: we can clearly see (fig.\ref{plan19juillet01} bottom) that $E_{g}$ eventually cancels out for a value of $r$ greater than $\simeq 0.13$. In other words, although it has only been provided with $13 \%$ of possible TDS, the network is able to generalize to the remaining $87\%$ of cases, those it has never learned.

The second observation is that this convergence is far from uniform: we observe fluctuations in the generalization error (fig.\ref{plan19juillet01} bottom) whose amplitude is large, being of the order of magnitude of $E_{g}$ averaged over the duration of the measurement. Similarly, the evolution of the number $N$ of hidden nodes with $r$ (fig.\ref{plan19juillet01} top) shows the same kind of fluctuations, with a start at $1$, several intermediate peaks including one at $34$ and a stabilization at $4$ after the grokking transition.
A priori, the non-uniformity of convergence is not really a surprise for a greedy algorithm, since it is generally impossible to demonstrate that local optimization (of the generalization) at one node is not at the expense of optimization at another node. But the high amplitude of the fluctuations is indicative of an avalanche dynamic, characterized by the existence of pitfalls into which the entire network plunges, and which it abruptly leaves only when it learns a new lesson.

\section{Statistics of grokking}
\label{Statistics}
In the following we discuss (always on the parity problem) the statistical aspects of learning. 

\subsection{Grokking as a statistical phenomenon}
\begin{figure}[!h]
\resizebox{0.40\textwidth}{!}{
\includegraphics[]{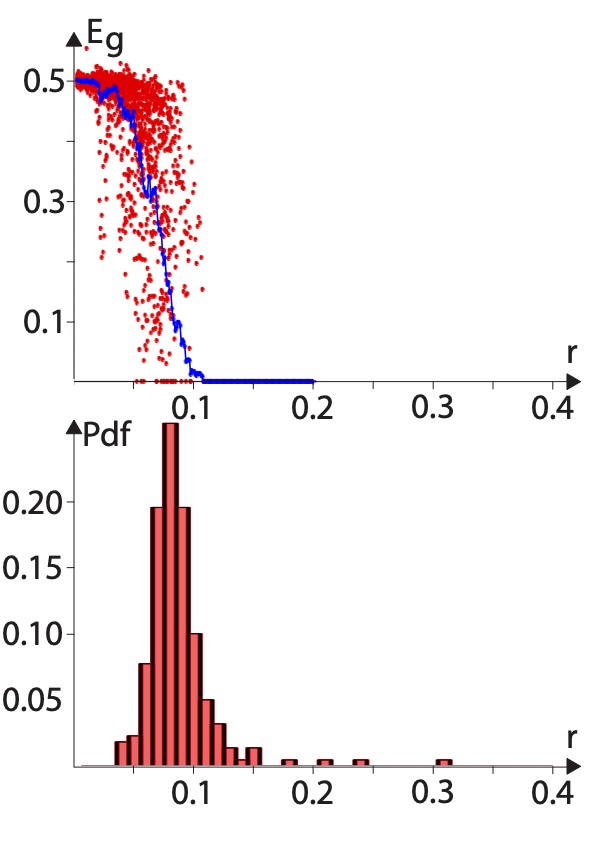}
}
\caption{Parity problem with $N_{e}=10$. Top: Evolution of the generalization error $E_{g}$ as a function of $r$ for $20$ independent training sets. The dots are the numerical measurements and the solid blue line represents the mean generalization error averaged over these $20$ training sets. Bottom: PDF of $r_{g}$  based on a sample of $220$ independent training sets.}
\label{plan19juillet02}
\end{figure}

As in fig.\ref{plan19juillet01} bottom, the top of fig.\ref{plan19juillet02} shows the evolution of the generalization error $E_{g}$ as a function of $r$ for the parity problem. The difference is that here we have superimposed the trajectories of 20 independent training sets. Clearly, $E_{g}$ always cancels out for a sufficiently large value of $r>r_{g}$, whatever the trajectory considered. On the other hand, $r_{g}$ undoubtedly depends on the trajectory considered. This is because the learning process, and therefore the value of the grokking threshold $r>r_{g}$, are stochastic phenomena in more ways than one. First, there are $2^{N_{e}}$ possible network entry codes and therefore ${(2^{N_{e}})!}\over{(2^{N_{e}}-m)!}$ ways of choosing $m$ codes from $2^{N_{e}}$, the order of choice being relevant. Some of these training sets are particularly un-pedagogical. For example, a bad teacher trying to explain the parity problem might start by presenting all the entry codes that will give a 0 answer, then all the others that will give a 1. He will obtain a “eureka” with difficulty from his pupil only for $r$ much greater than $1/2$. Fortunately, the number of training sets is gigantic ($>10^{300}$ for $N_{e}=10$ and $m=102$) and the probability of randomly drawing a wrong training set is very low. Second, even for a single training set, the algorithm includes many arbitrary choices: during the splitting cascade, the order of nodes splitting in two is stochastic. Also during conflict resolution, the choice of which pairs of nodes will merge is also random.

The threshold $r_{g}$ is therefore a stochastic variable.  The bottom of fig.\ref{plan19juillet02} is a numerical measure of its probability density function (PDF), performed on a (small) sample of $220$ independent training sets. The sample size is limited by the length of the calculation times. Nevertheless, in view of the narrowness of the probability distribution, it seems reasonable to define a mean value. At the top of Fig.\ref{plan19juillet02}, the continuous blue curve represents the mean generalization error as a function of $r$, $<E_{g}(r)>$, averaged over the $20$ independent learning trajectories represented by the points.

\subsection{Finite size effects}
Simple statistical physics models, such as single-layer perceptrons, predict cancellation of the generalization error $E_{g}$ with the number $m$ of TDS, when $m>= \alpha N_{p}$ where $\alpha \simeq 1. 39$ and $N_{p}$ is the number of independent adjustable parameters. $E_{g}$ tends gently towards $0$ when the adjustable parameters evolve continuously, and abruptly when they take only discrete values \cite{seung1992statistical,sompolinsky1990learning,gyorgyi1990first}.
Since our algorithm involves several layers, with a constantly varying number of nodes and connections (and for which $N_{p}$ remains to be defined), these theoretical results are not necessarily relevant. Nevertheless, we do observe a rapid fall in the generalization error with $r$, as expected for discrete models.
Also, numerous recent numerical observations on a heterogeneous set of algorithmic data report that the grokking threshold value $r_{g}$ decreases with the number of input interface nodes $N_{e}$ \cite{power2022grokking}. In \cite{vzunkovivc2022grokking,vzunkovivc2024grokking}, this same dependency is derived analytically for simple, solvable models.

\begin{figure}[!h]
\resizebox{0.40\textwidth}{!}{
\includegraphics[]{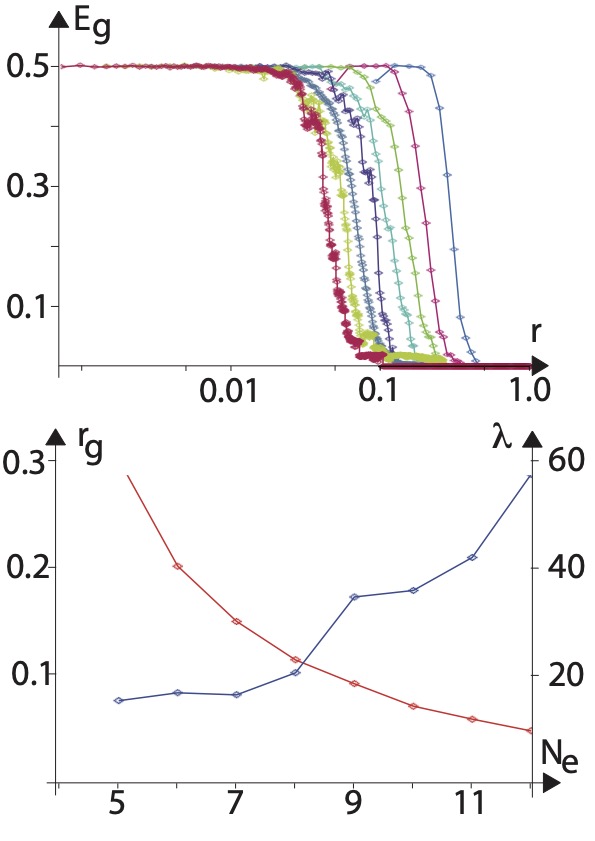}
}
\caption{Top: Plots of $E_{g}$ versus $r$ for several values of $N_{e}$. From right to left, we have $N_{e}=5,6,7,8,9,10,11,12$. Each curve is obtained by averaging over 20 training sets; the points are the numerical measurements, the solid lines linking them are merely a visual aid. Note the logarithmic scale of the abscissa.
Bottom: Plots of $r_{g}$ and $\lambda$ defined in eq.\ref{ansatzDuFit} versus $N_{e}$. We were only able to estimate the error bars correctly for the case $N_{e}=10$, for which the number of training sets is sufficiently large (because of the bottom of fig.\ref{plan19juillet02}). We then measure a relative error of $3\%$ for $r_{g}$ and $16\%$ for $\lambda$.} 
\label{plan19juillet03}
\end{figure}

Here, we're interested in the effect of the network input size on the grokking transition, but for networks which have been generated by our algorithm.  Fig.\ref{plan19juillet03} top plots the evolution of the generalization error as a function of $r$, for several sizes $N_{e}$ of the number of input interface nodes. Each curve is obtained by averaging over $20$ independent training sets.  The problem studied is always that of parity because it is a non-trivial example for which $r_{g}$ is small and allows reasonably large values of $N_{e}$ to be investigated, but the observations are identical for the other problems studied. We therefore observe not only that $r_{g}$ decreases with $N_{e}$, but also that the size of the zone over which $E_{g}$ falls seems to decrease with $N_{e}$.  To quantify these observations, we fit the curves in fig.\ref{plan19juillet03} top with an ansatz of the form:
\begin{equation}
E_{g}(r)={{1}\over{4}} \left[ 1-erf \left( \lambda \left(r-r_{g}\right) \right)\right]
\label{ansatzDuFit}
\end{equation}

where $\lambda^{-1}$ has the dimension of an interval of $r$ and $erf$ denotes the error function. Fig.\ref{plan19juillet03} bottom plots $r_{g}$ and $\lambda$ as a function of $N_{e}$ and confirms the previous observation: larger $N_e$ brings a sharper transition, which is reached for smaller values of $r_g$. Unfortunately, the asymptotic behavior at large $N_{e}$, potentially highly informative, is not accessible with our current computational resources.

\section{Examples of application}
\label{Examples}
Finally, we show that the proposed algorithm can indeed generate networks capable of performing a variety of simple or complex classification tasks.

These tasks have been chosen to be as heterogeneous as possible. Some are related to arithmetic operations. The $N_{e}$ input interface nodes are divided into 2 groups of respective sizes $N_{e1}$ and $N_{e2}$, with $N_{e1}+N_{e2}=N_{e}$. Each set of bits is then interpreted as the binary code of an integer $n_{1}$ (resp $n_{2}$). We then define $3$ operations:
\begin{enumerate}
    \item ADD: A priori, the binary encoding of the integer $n_{1}+n_{2}$ requires $L=max(N_{e1},N_{e2})+1$ bits. The ADD function returns the most significant bit, i.e. the one associated with $2^{L-1}$.
    \item MUL: Binary encoding of the integer $n_{1}*n_{2}$ requires at most $L=N_{e1}+N_{e2}$ bits. MUL returns the most significant bit, i.e. the one associated with $2^{L-1}$.
    \item SUP: returns 1 if $n_{1}>=n_{2}$, and $0$ otherwise.
\end{enumerate}
Other tasks are associated with the presence of a particular pattern in the binary code entered:
\begin{enumerate}
    \item PAR: returns $0$ if the number of input interface bits equal to $1$ is even, and $1$ otherwise.
    \item TRI: returns $1$ if there are at least $3$ equal successive input interface bits in the binary code, and $0$ otherwise.
\end{enumerate}
Finally, some tasks correspond to more complicated operations. Let $n$ be the integer associated with the binary coding provided by the $N_{e}$ input interface bits. Then:
\begin{enumerate}
    \item FIB: returns 1 if $n$ is part of the Fibonacci sequence ${1,1,2,3,5,8,13...}$, and $0$ otherwise.
    \item PRI: returns 1 if $n$ is a prime number, and $0$ otherwise.
\end{enumerate}
Alongside these examples, we also tested our algorithm on a stochastic classifier, RAN, returning $0$ or $1$ randomly, with probability $1/2$. For this problem, the training set can be learnt as perfectly as for the other problems but generalization is by construction impossible. We therefore expect a generalization error $E_{g}$ of the form 
\begin{equation}
E_{g}(r)={{1}\over{2}} \left(1-r\right)
\label{eqRAN}
\end{equation}

\begin{widetext}
\begin{figure*}[t]
\resizebox{1.00\textwidth}{!}{
\includegraphics[]{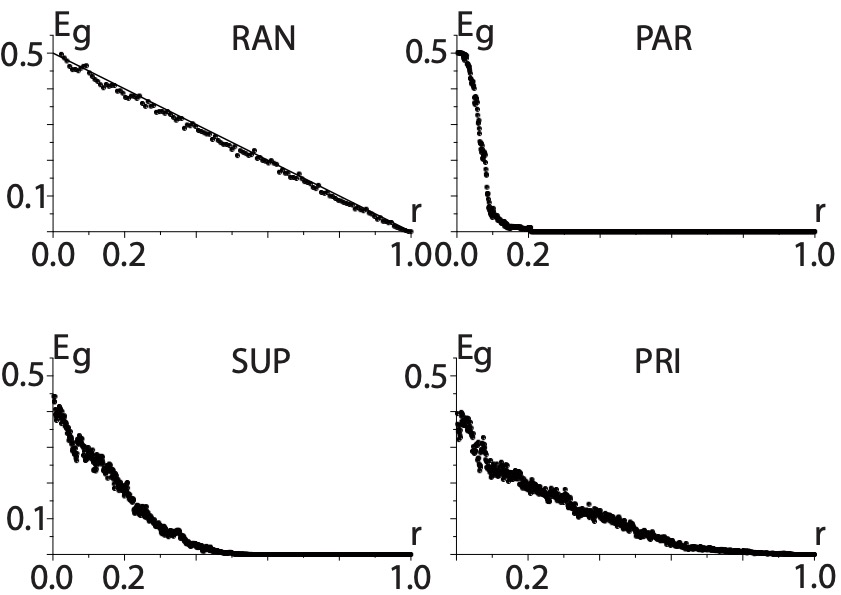}
\includegraphics[]{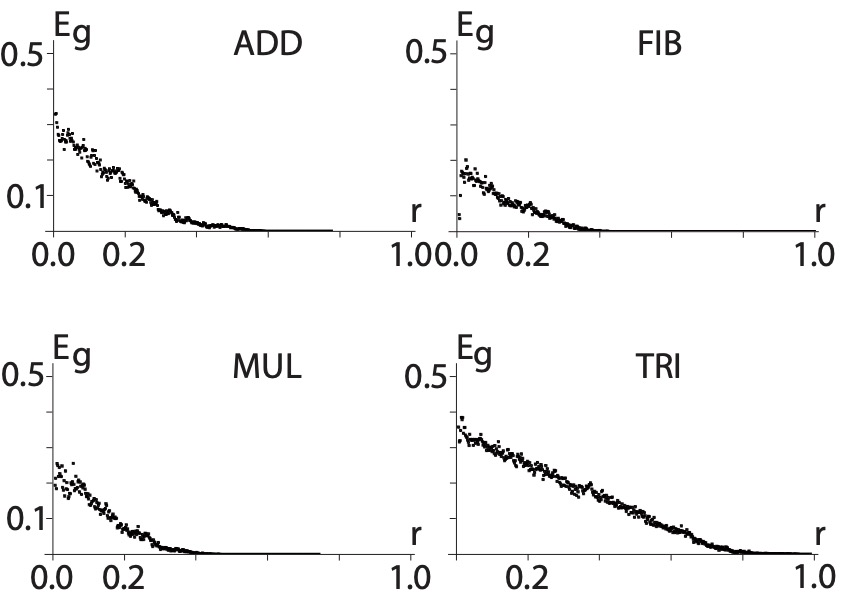}
}
\caption{Evolution of the generalization error $E_{g}$ as a function of $r$ for different problems with $N_{e}=9$. The title of the figure refers to the problem considered (see text). For RAN, the continuous straight line is the theoretical prediction of eq.\ref{eqRAN}.} 
\label{plan19juillet04}
\end{figure*}
\end{widetext}

Fig. \ref{plan19juillet04} shows the results for all the examples studied. The first observation is that our algorithm provides a network capable of generalizing in all cases where there is a rule to grok and only in those cases.
There are neither false negatives nor false positives. Another point to note is that the threshold of the grokking transition is highly dependent on the problem under consideration. So, alongside the theoretical classification of problem difficulty in terms of the size of the computing resources required (memory and number of instructions in a Turing machine), the numerical value of $r_{g}$ could be considered as an “experimental” value.

\section{Discussion}
\label{sec:discussion}

As we have shown above on several examples, the proposed learning approach can bring a network to grokking a variety of tasks. Beyond demonstrating that learning and generalizing can take place without a global cost function, these results may also have strong implications about generalization itself. 
Indeed, in order to answer the question of what distinguishes a neural network that generalizes well from one that generalizes poorly, a number of statistical learning theories have proposed measures of complexity capable of controlling generalization error \cite{bartlett2002rademacher,mukherjee2002statistical,bousquet2002stability,poggio2004general}. Nevertheless in \cite{zhang2021understanding}, the authors report on numerical results that "rule out the above complexity measures as possible explanations for the generalization performance of state-of-the-art neural networks".  Our study doesn't answer the question of the origin of the ability to generalize either, but it highlights 2 important points:  
1) it provides a theoretical framework for designing learning networks in which learning and generalization are completely decoupled. Whatever complexity measure is used to control the generalization error, it can now be studied without concern for learning mechanisms. 
2) The network's ability to generalize occurs as an emergent property, i.e. as the collective result of a local dynamic of expressiveness optimization at node level.

The key features of our approach can be summarized as follows:
\begin{enumerate}
	\item Standard neural networks (and also KAN networks \cite{liu2024kan}) optimize the edges of a fixed-topology network to minimize global error function, while our network stores information in LUTs and achieves generalization via dynamically evolving topology, distributing complexity both in nodes and edges.
    \item There is no global cost function to minimize, nor any gradient calculation. All operations are performed locally: (i) spatial locality, as the evolution rules involve only a single node or a pair of nodes, and (ii) temporal locality, as the temporal evolution takes place after each new learning. Our algorithm is therefore naturally parallelizable.
    \item  In a classic deep learning approach, the choice of meta-parameters (number of layers, number of nodes, number of connections) is more a matter of know-how than technique. Our algorithm solves this problem in an elegant way: all these choices are made naturally by the network itself as it evolves.
    \item Being able to decouple the problem of training data learning from that of seeking generalization greatly extends the range of possible ways of solving the later problems. The greedy algorithm is a simple solution, but more original (and certainly more efficient) procedures are conceivable.
\end{enumerate}

\section{Conclusion}
\label{Conclusion}
We have proposed and analyzed a new framework for neural computation based on dynamically evolving binary connections between dynamically evolving lookup tables. The learning phase does not rely on global minimization but instead on storing examples in LUTs before dynamically reconfiguring the network topology via a greedy algorithm which maximizes local expressivity. We have shown that such an evolved network can generalize many binary classification example tasks and that grokking shows distinctive features of a phase transtion. Beyond getting rid of a global cost function, our results also shed a different light on generalization, which can be sought independently from example learning, an observation which may contribute to rethinking generalization. 

The present study is only a proof of concept, limited to very small input codes ($N_{e}\le 12$), several orders of magnitude smaller than those commonly used by current learning machines. We are currently working on increasing this size by parallelizing the algorithm.

\bibliography{learning}
\bibliographystyle{apsrev4-1}

\end{document}